\documentclass[12pt]{amsart}
\usepackage{amsmath,amsfonts,amsthm,amssymb}

\newcommand\R{{\mathbb R}}

\renewcommand\S{\mathbb{S}}
\newcommand\supp{\text{ supp } }

 \newtheorem{theorem}{Theorem}
 \newtheorem{assumption}{Assumption A \!\!}

 \newtheorem{lemma}[theorem]{Lemma}
 \newtheorem{corollary}[theorem]{Corollary}
 
 \newtheorem{definition}[theorem]{Definition}
\newtheorem{proposition}[theorem]{Proposition}

\newcommand\e{\epsilon}
\renewcommand\d{\partial}

\newcommand\tchi{\tilde\chi}
\begin{document}

\title{Carleman estimates and absence of embedded eigenvalues}

\author{Herbert Koch}
\address{ Fachbereich Mathematik   \\ Universit\"at Dortmund }

\author{ Daniel Tataru}
\address {Department of Mathematics \\
 University of California, Berkeley}

\begin{abstract}
  Let $L = - \Delta - W$ be a Schr\"odinger operator with a potential
  $W\in L^{\frac{n+1}2}(\R^n)$, $n \geq 2$. We prove that there is no positive
  eigenvalue. The main tool is an $L^p-L^{p^\prime}$ Carleman type
  estimate, which implies that eigenfunctions to positive eigenvalues
  must be compactly supported.  The Carleman estimate builds on
  delicate dispersive estimates established in \cite{MR2094851}. We
  also consider extensions of the result to variable coefficient
  operators with long range and short range potentials and gradient
  potentials.

\end{abstract}

\maketitle

\section{Introduction}

Let $n \geq 2$. Suppose $W$ is a potential in $\R^n$ which decays at
infinity.  Then the Schr\"odinger operator
\[
 -\Delta_{\R^n} - W 
\]
has continuous spectrum $[0, \infty)$. In addition its spectrum may
contain eigenvalues which could be positive, negative of zero. 
 Positive eigenvalues in the continuous spectrum are
undesirable. They are very unstable since they are destroyed even by
weak interactions between the continuous spectrum and the eigenvalue
(see \cite{0917.35023}). Physically they correspond to trapped states
in the continuous spectrum, and they are difficult to handle
analytically. Moreover, excluding eigenvalues in the continuous
spectrum is often a first step toward scattering. There is an
extensive theory dealing with the absence of positive eigenvalues.

 It is well known that under weak  assumptions like 
\begin{equation}  
\lim_{|x| \to \infty} |x||W(x) | = 0  \label{decay} 
\end{equation} 
there are no positive eigenvalues. The argument uses 
Carleman estimates in three steps as follows. Suppose that 
\[ 
-\Delta u - Wu =  u  
\]
with $u \in L^2$, where the eigenvalue is normalized to $1$ by scaling. 
Then one proves that:

\begin{enumerate} 
\item \label{step1} The eigenfunction $u$ decays faster than polynomially at infinity.  
\item \label{step2} If $u$ vanishes faster than polynomially at
  infinity that $u$ has compact support.
\item \label{step3} If $u$ has compact support then it must vanish. 
\end{enumerate}

These arguments work for many Schr\"odinger operators.  However they
do not cover Schr\"odinger operators for several particles (which are
studied in \cite{MR85b:35038} and \cite{MR85g:35091}), neither do the standard 
arguments 
apply to the absence of bound states (i.e. $L^2$ solutions) in
nonlinear optics modeled by problems of the type
\[ -\Delta u = \omega u + a(x) |u|^\sigma  u  \]
with a bounded function $a$, because it is not clear how the assumption 
$u \in L^2(\R^n)$ is related to pointwise decay. 

On the other hand the assumption \eqref{decay} on pointwise decay 
is sharp: There is the famous Wigner-Von Neumann example of  a positive
 eigenvalue and a  potential decaying like $1/|x|$ but not better, see 
\cite{WN,MR58:12429c}.

Motivated by the above questions and by other potential applications 
one seeks to replace the pointwise bound \eqref{decay} by an $L^p$
bound. In terms of scaling any such bound must necessarily be weaker 
than \eqref{decay} due to  counterexamples by  Jerison and Ionescu
(\cite{MR2024415}) with potentials concentrated
close to $n-1$ dimensional planes. 
Jerison and Ionescu \cite{MR2024415} have recently obtained absence of
embedded eigenvalues for $W\in L^{n/2}$.
In this paper we  obtain the same result  for a larger class of
potentials which includes
 \begin{equation} 
 W \in L^{\frac{n+1}{2}}. \label{n+1}
\end{equation} 
We note that a higher index is better since it allows for potentials
with less decay at infinity. Another way to look at this is that such
a condition is mostly relevant for the low frequency part of $W$.  The
counterexample of Jerison and Ionescu (\cite{MR2024415}) shows that this 
is the highest possible exponent.

  Our method is robust enough so that it also allows us to add a long
 range potential, and also to replace the Laplacian with a (mildly)
asymptotically flat second order elliptic operator. The latter
generalization is more technical and less self-contained, so it is
discussed only in the last section.

Thus we consider potentials which are  the sum of weakly decaying  long
range potentials $V$ and short range potentials $W$.  We even include 
the eigenvalue  $\lambda>0$
into the long range potential and  study the problem 
\begin{equation} \label{fund}  
(- \Delta  - V ) u = W u. 
\end{equation}

To describe the long range potential we define the space $C^2_{\langle
  x \rangle}$ by
\begin{definition} \label{c2x} 
$C^2_{\langle x \rangle}$ is the space
  of $C^2_{loc}$ functions for which the following norm is finite:
\[
\Vert f \Vert_{C^2_{\langle x \rangle}} := \max\{ \sup_{x} | f(x)|,
\sup \langle x \rangle |Df|, \sup \langle x \rangle^2 |D^2 f|\}
 \] 
\end{definition} 

Then we introduce the condition

\begin{assumption}[The long range potential]
$V$ belongs to $ C^2_{\langle x \rangle}$ and satisfies
\begin{equation}  
\liminf_{|x|\to \infty} V > 0,  \qquad
\tau_0 := - \liminf_{|x|\to \infty} \frac{ x
    \cdot \nabla V}{4V}  < 1/2. 
\label{convex}
\end{equation}
\end{assumption}

The bound from below  on $V$ corresponds to the condition $\lambda >0$
while the last bound in \eqref{convex}  says that for large
$|x|$ the function $|x|^2$ is strictly convex along the null Hamilton flow 
for $-\Delta -V$, and thus guarantees nontrapping outside a compact
set.

To describe the short range potential we define the space
\begin{definition} \label{X} $X$ is the space of
  $W^{-\frac1{n+1},\frac{2(n+1)}{n+3}}_{loc}$ functions for which the
  following norm is finite:
\[
\| W\|_X =\sup_{u\in C^\infty_0} \Vert W u  
\Vert_{W^{-\frac1{n+1},\frac{2(n+1)}{n+3}}} /
  \Vert u  \Vert_{ W^{\frac1{n+1},\frac{2(n+1)}{n-1}}} \qquad n \geq 3
\]
\[
\| W\|_X =\sup_{u\in C^\infty_0} \Vert W u  
\Vert_{W^{-\frac1{3}+\epsilon,\frac65}} /
  \Vert u  \Vert_{ W^{\frac13-\epsilon,6}} \qquad n =2, \epsilon > 0
\]
  For a domain $D \subset \R^n$ we denote
\[
X(D) = \{ 1_D W;\ W \in X\}
\]
\end{definition}
Then we introduce 
\begin{assumption}[The short range potential] 
 $W$ belongs to $X_{loc}$ and 
can be decomposed as $ W=W_1+W_2$ where
\begin{equation}\label{w1} 
 \limsup_{j\to \infty} 
 \Vert W_1 \Vert_{X(\{ x| 2^j \le |x| \le 2^{j+1}\})} < \delta
\end{equation}
\begin{equation}\label{w2}
\limsup_{|x| \to \infty}  |x| |W_2(x)| < \delta.  
\end{equation}
\label{srp}\end{assumption}
The $W_2$ component corresponds to the $L^2$ Carleman estimates.
The class of allowed $W_1$ potentials includes $L^{\frac{n}2}$ and
$L^\frac{n+1}2$ or even better\footnote{$l^{\frac32}L^{1+}$ if $n=2$.}
 $l^\frac{n+1}2(L^{\frac{n}2})$
where the $l^\frac{n+1}2$ norm is taken with respect to a partition of
$\R^n$ into unit cubes.

Our main result is 
\begin{theorem} \label{uniqueness} Assume that  $V$ 
and $W$ satisfy
  Assumptions A1 and A2, let $\tau_1>\tau_0$ and assume  that $\delta$ 
is sufficiently small. Let $u \in
  H^1_{loc}(\R^n)$ satisfy \eqref{fund} and
  $(1+|x|^2)^{\tau_1-\frac12} u \in L^2$.
  Then $u\equiv 0$.
\end{theorem} 

By comparison, the result of Jerison and Ionescu \cite{MR2024415} applies
to the case $V =1$ and $W \in L^{\frac{n}2}$, $n\geq 3$. We note that the
exponent $p=n/2$ is critical for weak unique continuation; for smaller
exponents there are examples of compactly supported eigenfunctions,
see \cite{MR2002m:35020}.  

The conditions \eqref{w1} and \eqref{w2} have a different scaling
behavior. Nevertheless both are sharp, which can be seen by 
the Wigner-Von Neumann example  and the non radial counter example of
Jerison and Ionescu.

The proof uses Carleman estimates, following the same three steps
indicated above.  A combined $L^2$- $L^p$ Carleman inequality replaces
the previous $L^2$ Carleman inequalities.  Proving such inequalities
is a highly nontrivial task and relies on the bounds established in
\cite{MR2094851}.  Conjugation of the operator $-\Delta - V$ with the weight
of the Carleman inequality leads to a non-selfadjoint partial
differential equation. A pseudo-convexity type condition is satisfied,
but it degenerates for large $x$. This is related to the fact that the
anti-selfadjoint part of the conjugated operator decays for large $x$
in relevant coordinates.

Compared to earlier work and to the steps outlined above, we also
consider a different family of weights in the Carleman estimates.
Precisely, we begin with weights of the form $h(x) = e^{\tau
  \sqrt{|x|}}$ for part \ref{step2} of the argument, which we then
flatten at infinity for part \ref{step1}. This yields a more robust
argument, and also better results in the variable coefficient case.

The paper is organized as follows. In the next section we state all
the $L^p$ Carleman estimates and show how they lead to the result on
the absence of the embedded eigenvalues. 

There are two main ingredients in the proof of the $L^p$ Carleman
estimates. The first is the $L^2$ Carleman estimates, which are proved
in Section \ref{l2sect}. The second is a dispersive estimate for
second order operators which is obtained in Section \ref{MR2094851} using an
earlier result of the authors, namely Theorem 3 of \cite{MR2094851}. This is
of independent interest so we state it in more generality than needed
here.

 The $L^p$ estimates are proved Section \ref{dispersive}.  The
$L^2$ bounds obtained earlier are used to localize the $L^p$ bounds to
small spatial scales. Then we can rescale to a setting where the
general dispersive estimates of Theorem \ref{loc} apply.

Finally, in the last two sections we discuss the extension of the
results to second order elliptic operators with variable but
asymptotically flat coefficients as well as unbounded gradient
potentials. This goes along the same lines.

\section{Carleman estimates and embedded eigenvalues}

As explained above the proof depends on Carleman inequalities. In this
section we explain the Carleman inequalities and their application
whereas most of the proofs are postponed to the remaining sections. 

Let $1\le p \le \infty$ and $s\in \mathbb{R}$. We define the Sobolev space 
$W^{s,p}(\R^n)$ by the norm $\Vert f \Vert_{W^{s,p}} = \Vert  (1+|D|^2)^{s/2} f
\Vert_{L^p}$ and $W^{s,p}(U)$ for open subsets $U$ of $\R^n$ through its norm
which is the infimum of the norm of extensions. 

Given a measurable function $f$ and the Sobolev space $W^{s,q}$ 
we define the norm 
\[ 
\Vert f \Vert_{l^pW^{s,q}} 
= \left( \sum_{j=1}^\infty \Vert f \Vert^p_{W^{s,q}(\{
  2^{j-1} \le |x| \le 2^{j+1}\}) } \right)^{1/p} 
\]
with the obvious modification for $p= \infty$. 

Our Carleman estimates have the form
\begin{equation} \label{cm}
\begin{split}
 \Vert e^{h(\ln(|x|))}  &v \Vert_{l^2W^{\frac1{n+1}, \frac{2(n+1)}{n-1}}} 
+  \Vert e^{h(\ln(|x|))}  \rho  v \Vert_{L^2} 
\lesssim  \\ & \inf_{f_1+f_2=(-\Delta - V)v }  \Vert 
e^{h(\ln(|x|))} \rho^{-1}   f_1 \Vert_{L^2} + \Vert e^{h(\ln(|x|))}
f_2 \Vert_{l^2W^{-\frac{1}{n+1},\frac{2(n+1)}{n+3}}}
\end{split}
\end{equation}
where $\rho$ is given by
\begin{equation} \label{rho} 
 \rho =  \left( \frac{ h'(\ln(|x|))}{|x|^2} +  \frac{ h'(\ln(|x|))^2
   h''_+(\ln(|x|))}{|x|^4} \right)^\frac14
\end{equation}    
with $h''_+$ denoting the positive part of $h''$.
As a general rule, the function $h$ is chosen to be 

(a) increasing, $h'\geq \tau_0$, with $h'(0)$ large.

(b) slowly varying on the unit scale,  $|h^{(j)}| \lesssim h'$ for $j=2,3,4$.

(c) strictly convex for as long as $h'(\ln(|x|)) \gtrsim |x|$.

More precise choices are made later on for convenience, but 
the estimates are in effect true for all functions $h$ satisfying the
above conditions. 

The two terms in $\rho$ have different origins. The second one simply
measures the effect of the convexity of the function $h$.
The first one, on the other hand, is due to the presence of the long
range potential, which provides some extra pseudoconvexity for large $|x|$.

A simplifying assumption consistent with the choices of weights in
this paper is to strengthen (c) to 

(c)' $h''(\ln(|x|)) \approx h'(\ln(|x|))$  for as long as $h'(\ln(|x|)) \gtrsim |x|$.

This allows us to simplify the expression of $\rho$ to
\begin{equation} \label{rhoo} 
 \rho =  \left( \frac{ h'(\ln(|x|))}{|x|^2} \left (1+  \frac{
       h'(\ln(|x|))^2}{|x|^2} \right) \right)^\frac14
\end{equation}    

Our Carleman estimates use weights which grow exponentially,
but also allow for the possibility of leveling off the weight for
large enough $|x|$.

\begin{proposition} \label{carleman} Suppose that  $V$ satisfies
  Assumption A1. There is a universal constant $\varepsilon_0$ 
such that with 
\begin{equation}
h'_\epsilon (t)=  \tau_1 + (\tau e^\frac{t}2 -\tau_1)
\frac{\tau^2}{\tau^2 +\epsilon e^t}
\label{he}\end{equation}
 \eqref{cm} holds with $h = h_\epsilon$ for all $|\varepsilon|\le
 \varepsilon_0$, $v$ supported in
$|x| > 1$ and satisfying $ |x|^{\tau_1-\frac12}  v \in L^2$, uniformly
with respect to $\tau$ large enough.
\end{proposition}

The coefficient $\frac12$ in the exponent is chosen somewhat
arbitrarily. However, it must be smaller than $1$ in order for stage
(c) above to be reached. This is necessary if we are to be able to
taper off the weight at infinity.  We continue with a short discussion
of the weight $h_\epsilon$.

For small $t$ it is uniformly convex in the sense that $h_\epsilon'' \approx h_\epsilon'$.
The first interesting threshold for it is $t_0$ defined by
\[
e^{t_0} \approx \tau^2
\]
This implies that $h'_\epsilon(t_0) \approx e^{t_0}$.
In the range $[0,t_0]$ the last factor in \eqref{he} is largely irrelevant,
and $h'_\epsilon$ behaves like an exponential.
In this region, the pseudoconvexity in the Carleman estimates 
is  produced by the convexity of $h$.

After $t_0$ $h_\epsilon$ is still convex, roughly up to $t_1$ defined by   
\[
e^{t_1} \approx \epsilon^{-1} \tau^2
\]
The region $t_1+O(1)$ contains both the inflexion point $t_1$ and the
maximum point for $h'_\epsilon$.  In between $t_0$ and $t_1$ the
pseudo-convexity comes from the potential term, while the contribution
from the convexity of $h_\epsilon$ is still positive but smaller. 

Beyond $t_1+O(1)$ the function $h'_\epsilon (t)-  \tau_1$ decays 
in an exponential fashion. The last interesting threshold is $t_2$
where $h'_\epsilon$ approaches $1$, given by
\[
e^{t_2} \approx \epsilon^{-2} \tau^6
\]
Between $t_1$ and $t_2$ there is still convexity coming from the
potential $V$, which suffices in order to control the lack of
convexity of $h_\epsilon$. Finally, after $t_2$ the pseudoconvexity 
in the classical sense is lost, but there remains an Airy type gain 
to push the estimates through.

\begin{proof}[Proof of Theorem \ref{uniqueness}]
Here we show  that Proposition \ref{carleman} implies Theorem \ref{uniqueness}.

{\bf STEP 1:} We prove that $u$ decays at infinity faster than
$e^{-\tau \sqrt{|x|}}$.  We choose $R$ large enough so that (see
Assumption A2)

\begin{equation}\label{w11}  \sup_{2^{j+1} > R} 
 \Vert W_1 \Vert_{X(\{ x| 2^j \le |x| \le 2^{j+1}\})} < 2\delta
\end{equation}
\begin{equation}\label{w22}
\sup_{|x| > R}  |x| |W_2(x)| < 2\delta.  
\end{equation}
Choose $\phi \in C^\infty$ be identically $1$ for
$|x| \ge 2R$ and  $0$ for $|x| \le R$. We set $v=\phi u$.  Then
\[ 
-\Delta v - Vv = W v   - (\Delta \phi)u - 2 \nabla \phi \cdot
\nabla u   
\]
For $\tau_1$ as in Theorem \ref{uniqueness} we have $
|x|^{\tau_1-\frac12} v \in L^2$, therefore we can apply Proposition
\ref{carleman} with $\epsilon > 0$ to $v$ to obtain 
\[
\begin{split} 
 \Vert e^{h_\epsilon(\ln|x|)}  v \Vert_{l^2W^{\frac1{n+1},\frac{2(n+1)}{n-1}}} 
&\! +\!  \Vert e^{h_\epsilon(\ln|x|)}\rho  v \Vert_{L^2}  \lesssim 
 \Vert e^{h_\epsilon(\ln|x|)} \rho^{-1} (|u| +
|\nabla u|) \Vert_{L^2(B_{2R} \backslash B_R)}\\
  & + \Vert e^{h_\epsilon(\ln|x|)}  W_1 v \Vert_{l^2 W^{-\frac1{n+1}, \frac{2(n+1)}{n+3}}}
+  \Vert e^{h_\epsilon(\ln|x|)} \rho^{-1}  W_2 v \Vert_{L^2}
\end{split}
\] 
By \eqref{w11}, \eqref{w22} if $\delta$ is small enough then we can
absorb the last two right hand side terms on the left to obtain
\[
 \Vert e^{h_\epsilon(\ln|x|)}  v \Vert_{l^2W^{\frac1{n+1},\frac{2(n+1)}{n-1}}} 
 +  \Vert e^{h_\epsilon(\ln|x|)}\rho  v \Vert_{L^2}  \lesssim 
 \Vert e^{h_\epsilon(\ln|x|)} \rho^{-1} (|u| +
|\nabla u|) \Vert_{L^2(B_{2R} \backslash B_R)}
\]
Then letting $\e \to 0$  in the definition of $h$ yields 
\begin{equation} 
 \Vert e^{\tau \sqrt{|x|}}  v \Vert_{l^2W^{\frac1{n+1},\frac{2(n+1)}{n-1}}} 
+ \Vert e^{\tau \sqrt{|x|}} \rho   v \Vert_{L^2}
\lesssim \Vert  e^{\tau \sqrt{|x|}}  \rho^{-1} (|u| +
|\nabla u|) \Vert_{L^2(B_{2R}\backslash B_R)}.
\label{lt}\end{equation}
which shows that $v$ and therefore $u$ is rapidly decaying at infinity.

{\bf STEP 2:} We prove that $u$ vanishes outside a compact set.
This is done using \eqref{lt} (which can also be derived directly from
Proposition~\ref{carleman} as above). From \eqref{lt} we obtain
\[
R^{-1} e^{-\tau \sqrt{2R}} \Vert e^{\tau \sqrt{|x|}} v
\Vert_{l^2W^{\frac1{n+1},\frac{2(n+1)}{n-1}}} + \Vert e^{\tau
  \sqrt{|x|}} \rho v \Vert_{L^2} \lesssim \Vert (|u| + |\nabla u|)
\Vert_{L^2(B_{2R}\backslash B_R)}.
\]
Letting $\tau \to \infty$ shows that $v=0$ outside $B_{2R}$. Then the  same
holds for $u$.

{\bf STEP 3:} We prove that $u$ is identically $0$.  Assume by
contradiction that this is not the case, and choose $r$ minimal so
that $u$ is supported in $\overline{B(0,r)}$.  Our problem is scale invariant, so
without any restriction in generality we can assume that $r> 1$.
Take $x_0 \in \supp u$ with $|x_0| = r$. The problem is also invariant
with respect to translations so we can assume instead that
$\supp u \in \overline{B(x_0,r)}$ and $2x_0 \in \supp u$.

To reach a contradiction we prove that there is
$\alpha > 0$ so that  $u$ is supported in $B(0,2r-\alpha)$.
 This follows as in STEP 2
provided we know that for every $\delta >0$ we can find $\rho >0$ such 
that 
\[ \|W_1 v \|_{W^{-\frac1{n+1},\frac{n+1}{n+3}} }
\le \delta \Vert v  
\Vert_{W^{\frac1{n+1},\frac{n+1}{n-1}}}, 
\qquad \supp v \subset  B(2x_0,\rho)
\]
Then $\alpha$ is chosen so that 
\[
\{ 2r-\alpha < |x| <  2r\} \cap  B(x_0,r) \subset B(2x_0,\rho)
\]  
 Due to our
choice of $W$ this is a somewhat technical matter which is left for
Proposition~\ref{small} in the appendix. This step can be approached 
alternatively by the unique continuation results of \cite{MR2094851}. 
\end{proof}

\section{The $L^2$ Carleman estimates}

\label{l2sect}

In this section we obtain the $L^2$ Carleman inequalities. 

\begin{proposition} \label{l2} 
  Suppose that $V$ satisfies Assumption A1. Let $h$ be as in
  \eqref{he} and $\rho$ as in \eqref{rho}.  Then for all $u$
  satisfying $ |x|^{\tau_1-\frac12}  u \in L^2$ we have
\begin{equation}
  \Vert e^{h(\ln|x|)} \rho  u \Vert_{L^2} 
+   \Big\Vert \frac{|x|}{ h'(\ln |x|)+|x|} e^{h(\ln|x|)} \rho  \nabla u \Big\Vert_{L^2}   \lesssim 
 \Vert 
e^{h(\ln|x|) } \rho^{-1}   (\Delta + V) u  \Vert_{L^2}. 
\label{rhoh} \end{equation}
uniformly with respect to $\tau$ sufficiently large and $0 < \varepsilon \leq \varepsilon_0$.
\end{proposition} 

\begin{proof}
We use a conformal change of coordinates
\[
 t = \ln |x|, \qquad    y = x/|x| \in \mathbb{S}^{n-1}  
\] 
Denote 
\[
\Delta u = g
\]
and set
\[ 
v(t,y) = e^{(n-2)t/2}  u(e^ty),  \quad f(t,y) = e^{(n+2)t/2} g(e^t y)  
\]
 A routine computation shows that
\[  
|x|^{(n+2)/2}(\Delta  + V)|x|^{(n+2)/2}   =  \frac{\d^2}{\d t^2} 
+ \Delta_{\S^{n-1}}- ( (n-2)/2)^2
\]
therefore $v$ solves the equation
\begin{equation}\label{fund3} 
Lv = f, \qquad L =  \d_{t}^2 + \Delta_{\S^{n-1}}  - \left((n-2)/2\right)^2   + e^{2t} V
\end{equation}
We also note that part of Assumption A1 in the new coordinates we get
\[
- \liminf_{t \to \infty} \frac{V_t}{4V} = \tau_0 < \frac12
\]
 By \eqref{convex} we slightly readjust $\tau_0$ and  choose $t_0$
so that
\begin{equation}
-\frac{V_t}{4V} \leq \tau_0 < \frac12, \qquad t > t_0
\label{vta}\end{equation}

For any exponential weight  $h$ we have
\begin{equation}
 \int e^{2h(\ln |x| ) }   |u|^2 dx =  \int_\R \int_{\mathbb{S}^n} 
e^{2h(t)+nt} |u(t y)|^2 \, dt \, dy  =  \Vert e^{h(t)}e^{t} v  
\Vert_{L^2(\R \times S^{n-1})}^2, 
\end{equation}
\begin{equation}
 \int e^{2h(\ln |x| ) }   |g|^2 dx =  \int_\R \int_{\mathbb{S}^n} 
e^{2h(t)+nt} |g(t y)|^2 \, dt \, dy  =  \Vert e^{h(t)}e^{-t} f  \Vert_{L^2(\R \times S^{n-1})}^2.
\end{equation}
Hence, in the new coordinates the bound \eqref{rhoh}
becomes
\begin{equation} 
\label{transformed} 
\| e^{h(t)} \rho_1  v\|_{L^2} +  \|  e^{h(t)} \frac{\rho_1}{e^t + h'(t)}\nabla v\|_{L^2}
         \lesssim \|e^{h(t)}  \rho_1^{-1}  f\|_{L^2},
\end{equation}
where $\nabla v$ is the gradient of $v$ with respect
to $y$ and $t$ and, by \eqref{rhoo},
\[
 \rho_1(t) = e^t \rho = h'(t)\left(e^{2t} + h'(t)^2\right)^\frac14
\]

To prove the above bound one would like to follow a standard
strategy. This means conjugating the operator with respect to the
exponential weight, and producing a commutator estimate for the
self-adjoint and the skew-adjoint part of the conjugated operator.
There are two small problems with this approach, both of which
occur in the region where $h'(t)$ is small. 

First we want to incorporate the weight $\rho_1^{-1}$ on the right,
which would require an additional conjugation. Where $h'$ is small
this cannot be treated as a small perturbation, so we really have 
to include $\rho^{-1}$ in the exponential weight.

This leads to a second difficulty. After including $\rho^{-1}$ in the
exponential weight the commutator between the self-adjoint and the
skew-adjoint part of the conjugated operator is no longer fully
positive definite and we need a slightly modified argument.

To handle both issues we prove a slightly more general result
and then we obtain \eqref{transformed} as a special case of it.
Precisely, we consider an exponential weight $\phi$ as follows:

(i) $\phi' \geq \tau_1 - \frac12$, and $\phi'(0)$ is large.

(ii) $1+\phi'$ is slowly varying on the unit scale, i.e.
\[
|\phi^{(j)}(t)| \lesssim 1+\phi'(t) \qquad j=2,3
\]

(iii) $\phi'$ can only have a limited exponential growth rate, 
$\phi'' \lesssim \frac34(1+\phi')$.

Together with (i) this yields the existence of a unique $t_0$
so that $\phi'(t_0)=e^{t_0}$. Our last assumption asks for uniform
convexity up to $t_0$:

(iv) $\phi''(t) \approx \phi'(t)$ for $0 \leq t \leq t_0+C$ for some
large parameter $C$.

We summarize the bound for the weight $e^\phi$:

\begin{lemma}
  Consider a weight function $\phi$ satisfying the conditions (i)-(iv)
  above.  Then for all $v$  which are
  supported in $t > 0$ and with $e^{\phi(t) + t}
  v \in L^2$ we have
\begin{equation}
  \Vert  e^{\phi(t)}  (e^{2t}+\phi'(t)^2)^\frac12  v \Vert_{L^2}  + \Vert  e^{\phi(t)}  \nabla  v
  \Vert_{L^2}  
\lesssim 
 \Vert  e^{\phi(t)}  (1+\phi')^{-\frac12}  L v \Vert_{L^2}.
\label{rhohh}\end{equation} 
\label{smalltau} \end{lemma}
\begin{proof}
First we conjugate with respect to the exponential weight.
If we set $w = e^{\phi(t)}v$ then $w$ solves the equation
\[
L_\phi  w =  e^{\phi(t) }  f, \qquad L_\phi = e^{\phi(t)} L e^{-\phi(t)}
\]
We decompose $L_h$ into a selfadjoint and a skewadjoint part,
\[
L_\phi^r = \partial_t^2 + \Delta - (\frac{n-2}2)^2 + e^{2t } V +
\phi'^2, \qquad 
L_\phi^i =  -\phi'\partial_t - \partial_t \phi' 
\]
The bound to prove is
\begin{equation}
  \Vert  (e^{2t}+\phi'(t)^2)^\frac12 w \Vert_{L^2}  + \Vert   \nabla  w
  \Vert_{L^2}  
\lesssim 
 \Vert   (1+\phi')^{-\frac12} L_\phi w \Vert_{L^2}.
\label{rhohhh}
\end{equation} 
The proof of this inequality is based on several integrations by
parts.  In a standard manner one verifies that the integrations by
parts below are valid if $e^{\phi+t} v \in L^2$.

We multiply $P_\phi w$ by $- \frac12 w_t$ 
and integrate by parts to obtain
\begin{equation} 
 \begin{split}
 \int \phi' |w_t|^2 \,  dy\, dt  & + 
  \int (\frac14 \phi'''+ \frac12 \phi'\phi'' )
 |w|^2 dt dy  + \int \frac{e^{2t }}4(2V+
V_t) w^2 \, dy \,  dt 
\\ = &  \frac12\int w_t  L_\phi w \,  dy\,  dt
\end{split} 
\end{equation} 
This computation is essentially like taking the commutator of
$L_\phi^r$ and $L_\phi^i$.   On the left we have mostly positive
contributions, with the following qualifications:

-the first term can be negative where $\phi' < 0$

-the $\phi'\phi''$ term can also be negative, but only for $t > t_0+C$
where it is controlled by the $V$ term.

-the $\phi'''$ term is controlled either by the $V$ term or by the
$\phi'\phi''$ term.

To correct the first term in the region where $\phi'$ is negative we
consider a cutoff function $\chi$ which equals $\delta$ in  $\{\phi' > 2\}$ and
which equals $1$ in $\{\phi' < 1\}$. Here $\delta$ is a small universal parameter
which we shall choose below.
Since $\phi'+1$ is slowly varying we
can assume that $\chi$ has uniformly bounded derivatives.

Multiplying $P_\phi w$ by $\chi^2(t) w$ and integrating gives
\begin{equation} 
\begin{split}
\| \chi w_t\|^2_{L^2} + & \| \chi \nabla w\|^2_{L^2} +
(\frac{n-2}2)^2 \|\chi w\|_{L^2}^2
-\int \chi^2 (e^{2 t } V+\phi'^2) |w|^2 \,  dy \, dt =\\ 
& \int \frac12 (\partial_t^2 \chi^2) w^2\,  dy\, dt +
 \int w L_\phi w \,  dy\,  dt.
\end{split}
\end{equation} 
We multiply this by  $\mu$ and add to the previous relation. This yields
\begin{equation}  
\begin{split} 
 \mu \| \chi \nabla w\|^2_{L^2} + & \int (\chi^2 \mu+\phi') |w_t|^2 \, dy \, dt + 
\int \left(\frac12 - \chi^2 \mu + \frac{V_t}{4V}\right)  e^{2t }V w^2\,  dy\,
dt  \\ +  &
\int( \frac12 \phi' \phi'' - \chi^2 \mu \phi'^2)  |w|^2\,  dy\,  dt \\ 
 = &  
\int  \big[- \phi'''/4  - (\frac{n-2}2)^2\mu \chi^2 + \d_{tt}^2 \chi^2 \mu
\big]    |w|^2 \, dy\,  dt 
\\ & +
\int (\chi^2 \mu w + \frac12 w_t)  L_\phi w\,  dy\,  dt. 
\end{split}
\end{equation} 
To ensure that the left hand side is positive definite 
we recall that for large $t$ 
\[
-   \frac{V_t}{4V} \leq  \tau_0 <  \tau_1 \leq
 \frac12 + \phi' 
\]
Hence if we choose $\mu$ positive so that
\[
\frac12 - \tau_1 < \mu < \frac12 - \tau_0
\]
 then the first three terms are positive definite.

For the fourth term we consider two possibilities. If $t < t_0+C$ then
$\chi = \delta$ while $\phi'' \approx \phi'$ so it yields a positive
contribution.
We choose the universal constant $\delta$ so that 
\[ \frac12 \phi' \phi'' - \chi \mu {\phi'}^2 \ge \frac14 \phi' \phi'' \]
if $t \le t_0+C$.  
 For larger $t$ this fourth integrand may be negative but then it 
is controlled by the third. The first term on the right hand side 
is controlled by the left hand side
and we obtain
\[
 \| \nabla w\|^2_{L^2} + \|(1+\phi')^\frac12 w_t\|_{L^2}^2 +
 \|(\phi'(t)^2 +e^{2t})^\frac12  w\|_{L^2}^2 \lesssim
 \int (\mu w + \frac12 w_t) L_\phi w dy dt
\]
The proof is completed by an application of the Cauchy-Schwarz
to the  right hand side. 

\end{proof}

{\em Proof of Proposition~\ref{l2}, continued.}

We obtain \eqref{transformed} from Lemma~\ref{smalltau}. For this we
need to associate to each weight $h$ a function $\phi$ satisfying
(i)-(iv) with the property that
\[
1+\phi' \approx h', \qquad (1+\phi')^{-\frac14} e^\phi \approx e^{h}
(h'^2 +e^{2t})^{-\frac14}
\]
 The natural choice for $\phi$ is
\[
\phi(t) = h(t) -\frac{t}2 + \frac14 \ln(1+ h'(t)) - \frac14 \ln(1+
e^{-t} h'(t))
\]
Then 
\[
\phi' = h'-\frac12 + \frac{h''}{4(1+h')}+ \frac{(h' -h'')
  e^{-t}}{4(1+e^{-t} h')}
\]
We verify the properties of $\phi$. It is easy to see that $1+ \phi'$ is
slowly varying. This implies that the last two terms in $\phi'$ are bounded and
have bounded derivatives. Hence the properties (ii)-(iv) follow 
from the similar properties of $h'$.

 It remains to check the bound $\phi' > \tau_1-\frac12$.
This is  clear when $h' \gg 1$
which corresponds to $\e e^{\frac{t}{2}} \ll \tau^3$. For larger $t$
we have
\[
h'(t) = \tau_1 + \frac1{\epsilon} \tau^3 e^{-\frac{t}2} (1+O(\tau^{-1}))
\]
and 
\[
h''(t) =- \frac1{2\epsilon} \tau^3 e^{-\frac{t}2} (1+O(\tau^{-1}) )
\]
Then
\[
\phi'(t) > \tau_1 - \frac12 +  \frac1{2\epsilon} \tau^3 e^{-\frac{t}2} (1+O(\tau^{-1}))
\]
so the desired bound is again verified. We note that what happens 
when $h'$ is small is not so important anyway; in this region we can
simply choose $\phi(t) = h(t) -\frac{t}2$.  \end{proof}

\section{A general dispersive estimate for second 
order operators} 
\label{MR2094851} 

In this section we study the second order operator\footnote{We use the
  summation convention here and in the sequel.}
\[
  L_\mu  =  \d_i a^{ij}(x) \d_{j}   + \mu^2 c(x)   - i \mu 
 ( b_j(x)\d^j + \d_j b^j(x) ),   
\]
in the unit ball $B \subset \R^n$, $n \geq 2$ with real coefficients
$a^{ij}$ and complex coefficients $b^j$ and $c$. Here $\mu$ is
sufficiently large and plays the role of a semiclassical parameter.
Concerning the type and regularity of the coefficients we assume that
\[
(REG) \quad \left\{ \begin{array}{l}
\text{the matrix $(a^{ij}(x))$ is real,
symmetric and positive definite} \\
\text{the functions $a^{ij}$,
$b^i$ and $c$ are  of class $C^2$}
\end{array}\right.
\]

 We define the symbol 
\[
l(x,\xi) = - \xi_i  a^{ij}(x) \xi_{j} +  c(x) + 2 b_j \xi_j
\]
The real part of $l$ is a second degree polynomial in $\xi$
with characteristic set
\[
{\text char}_x \Re l(x,\xi) = \{ \xi \in \R^n; \  \Re l(x,\xi) = 0 \}
\]
The geometric assumption on the operator $L$ is
\[
(GEOM) \quad \left\{ \begin{array}{l}
\text{for each $x$ the characteristic set ${\text char}_x \Re l(x,\xi)$} \\
\text{is an ellipsoid of size $\approx 1$.}
\end{array}\right.
\]

Our third hypothesis is concerned with the size of the Poisson 
bracket of the real and imaginary part of $L$. We are interested
in a  principal normality type condition of the form
\begin{equation}
|\{ \Re l(x,\xi), \Im l(x,\xi)\}| \lesssim \delta + |\Re l(x,\xi)| +
|\Im l(x,\xi)|
\label{pn}\end{equation}
where the relevant range for $\delta$ is $\mu^{-1} < \delta \ll 1$.
This would suffice for our purposes if in addition we knew that all
the coefficients of $l$ are of class $C^3$.  In general for technical
reasons we need to replace the inequality with a decomposition
\begin{equation}
\{ \Re l, \Im l\}(x,\xi) =  \delta q_0(x,\xi) + q_1^r(x,\xi) \Re l(x,\xi) +
q_1^i(x,\xi) \Im l(x,\xi) + q_2(x,\xi)
\label{pn1}\end{equation}
Thus our last assumption has the form

\[
(PN) \!\quad \left\{ \begin{array}{l} \text{the Poisson bracket $\{
      \Re l, \Im l\}$ admits a
      representation \eqref{pn1} where } \\ \\
    |\partial_x^\alpha \partial_\xi^\beta q_i(x,\xi)| \leq
    c_{\alpha\beta} \qquad |\alpha|
    \leq i \\ \\
    |q_0| \lesssim 1, \qquad |q_1^r| + |q_1^i| \lesssim 1, \qquad
    |q_2| \lesssim |l|
\end{array}\right.
\]

For $L$ in the class of operators described above we are interested in
constructing a parametrix $T$ which has good $L^{p'} \to L^p$ and $L^2
\to L^p$ mapping properties, while the errors are always measured in
$L^2$. A dual form of this also allows us to estimate the $L^p$ norm
of a function $u$ in terms of the $L^2$ norms of $u$ and $Lu$.

In the context of the Carleman estimates such parametrices allow us
to superimpose local $L^{p'} \to L^p$ bounds on top of the global $L^2
\to L^2$ estimates in order to obtain a global $L^{p'} \to L^p$ bound.

Such estimates are dispersive in nature and are strongly related to
the spreading of singularities in the parametrix $T$. This in turn is
determined by the nonvanishing curvatures of the characteristic set
${\text char}_x \Re l(x,\xi)$.

If $L$ has constant coefficients and real symbol then the theorem
below is nothing but a reformulation of the restriction theorem. If
$L$ has real symbol but variable coefficients then we are close to the
spectral projection estimates of C. Sogge \cite{Sbook}. In the case
when $L$ has constant coefficients but complex symbol some bounds
of this type were obtained in \cite{MR88d:35037}.

In the more general case considered here we rely on bounds and
parametrix constructions in the author's earlier paper \cite{MR2094851}. These
apply to principally normal operators. The operator $L_\mu$ is
principally normal on the unit spatial scale only if $\delta \approx
\mu^{-1}$. Otherwise, we use a better spatial localization to the
$(\delta\mu)^{-\frac12}$ scale. On one hand  $L_\mu$ is
principally normal on this scale, while on the other hand 
this localization is compatible with the $L^2$ estimates and this
allows us to easily put the pieces back together.

All Sobolev norms in the theorem below are flattened at frequency
$\mu$ instead of frequency $1$ as usual. Hence we introduce the 
notation
\[
W^{s,p}_\mu = \{ u \in S'; \ (\mu^2+D^2)^{\frac{s}2} u \in L^p\}
\]
with the corresponding norm. 

We note that the operator $L$ is elliptic at frequencies larger than
$\mu$ so all the estimates are trivial in that case. All the
interesting action takes place at frequency $\lesssim \mu$, where we
can identify all Sobolev norms with $L^p$ norms.

\begin{theorem} \label{loc}
Suppose that the operator $L_\mu$ satisfies the conditions (REG),
(GEOM) and (PN) 
 for some $\delta > \mu^{-1}$.
 Let $\phi \in C(B_2(0))$ have compact support. Then

A) There exists  an operator $T$ such that 
\begin{equation} \label{T1}
\begin{split}
\Vert Tf \Vert_{W^{\frac1{n+1}, \frac{2(n+1)}{n-1}}_\mu} &
   + (\delta \mu)^{1/4} \mu^{-1/2} \Vert Tf \Vert_{H^1_\mu} 
\\ &\lesssim   \inf_{f=f_1+f_2}
(\delta \mu)^{-1/4}  \mu^{-1/2} \Vert f_1 \Vert_{L^2} 
+\Vert f_2 \Vert_{W^{-\frac1{n+1}, \frac{2(n+1)}{n+3}}_\mu}
\end{split}
\end{equation}  
and 
\begin{equation}
\begin{split}
  \label{T2} 
(\delta \mu)^{-1/4}  \mu^{-1/2}  \Vert  & L T\phi f- \phi f  \Vert_{L^2}  \lesssim
\\&     \inf_{f=f_1+f_2}(\delta \mu)^{-1/4} \mu^{-1/2} \Vert f_1 \Vert_{L^2} 
 +\Vert f_2 \Vert_{W^{-\frac1{n+1}, \frac{2(n+1)}{n+3}}_\mu}
\end{split}
\end{equation} 
B) For all functions $u$  in $B_2(0)$ we have
\begin{equation}\label{Eu}
\begin{split}
\Vert \phi u \Vert_{W^{\frac1{n+1},\frac{2(n+1)}{n-1}}_\mu}
  & \lesssim   (\delta \mu)^{1/4} \mu^{1/2} \Vert u \Vert_{L^2} 
\\ &  + \inf_{Lu=f_1+f_2} (\delta \mu)^{-1/4} \mu^{-1/2} \Vert  f_1  \Vert_{L^2} 
 +\Vert f_2 \Vert_{W^{-\frac1{n+1}, \frac{2(n+1)}{n+3}}_\mu}
\end{split}
\end{equation} 
C) Suppose that in  addition the problem is pseudoconvex in the sense
that
\begin{equation}   
q_0(x,\xi)  \approx \delta \gg \mu^{-1} \qquad  x \in B_2(0), \ \ \mu
\gg 1
 \end{equation}
Then for all functions $u$ with  compact support in $B_2(0)$ we have 
\begin{equation}
\begin{split} \label{Eu2}
\Vert u \Vert_{W^{\frac1{n+1}, \frac{2(n+1)}{n-1}}_\mu} 
& + (\delta \mu)^{1/4} \mu^{1/2}  \Vert u \Vert_{L^2}  
    \\ & \lesssim  \inf_{Lu=f_1+f_2+f_3}
(\delta \mu)^{-1/4}  \mu^{-1/2} \Vert  f_1  \Vert_{L^2} 
  +\Vert f_2 \Vert_{W^{-\frac1{n+1}, \frac{2(n+1)}{n+3}}_\mu}
\end{split}
\end{equation} 
\end{theorem} 

The difficult part of this theorem is the existence of the rough 
parametrix in Part A.  This existence will be derived from  Theorem 3
in \cite{MR2094851}. The arguments repeat partially those of Section 3, 7 and
8 of \cite{MR2094851}.  

\begin{proof} 
  
  {\bf Part A.}  {\bf (i) Localization}.  We first reduce the problem
  to the case when $\delta = \mu^{-1}$.  This is done by localization
  to a small spatial scale and then by rescaling.  The appropriate
  spatial scale is $r = (\mu \delta)^{-\frac12}$.  We cover the
  support of $\phi$ with balls $B_j$ of radius $r$ and choose a
  subordinate partition of unity of the form
\[ 
\sum \phi_j^2 = 1 
\]
Suppose that within $B_j$ there exists a parametrix $T_j$ satisfying 
the desired estimates. Then we set 
\[ 
T = \sum_{j=1}^N \phi_j T_j \phi_j. 
\]
The bound \eqref{T1} for $T$ follows directly by square summing the 
similar bounds for $T_j$. For \eqref{T2} we compute
\[
I - L T = \sum_{j=1}^N \phi_j (I-LT_j)\phi_j +  \sum_{j=1}^N [L,\phi_j]
T_j \phi_j 
\]
For the first term we use \eqref{T2} for $T_j$ while for the second 
we estimate the commutators using \eqref{T1} for $T_j$.

 In order to obtain the localized parametrices $T_j$ we rescale $B_j$ 
to the unit scale. Then the problem reduces to the original one but 
with $\delta = \mu^{-1}$.

{ \bf (ii) The elliptic high frequency parametrix}.

For each $x$ the zero set of $\Re l$ is an ellipse contained in a
ball of radius $B_{R\mu}(0)$ with $R \sim 1$.  Let $\psi \in
C^\infty(\R^n)$ be a nonnegative radial radially decreasing function
supported in $B_2(0)$ and identically $1$ in $B_1(0)$.  Let $\phi$ be
as in the statement of the theorem. We fix a nonnegative function
$\phi_0 \in C^\infty(B_2(0))$, identically $1$ on the support of
$\phi$.  We define $T_{high}$ by its Weyl symbol
\[ 
\phi_0(x)  l_\mu^{-1} (x,\xi) (1-\psi(\xi/\mu R)) \phi_0(x). 
\] 
Then the following $L^2$ bounds are immediate:
\[ 
\Vert T_{high} f\Vert_{H^1_\mu}  
  \lesssim   \Vert f \Vert_{H^{-1}_\mu} 
\]
\[ 
\Vert (1-LT_{high}) (1- \psi(D/(2\mu R)))   \phi f \Vert_{L^2}  
  \lesssim  
 \Vert f \Vert_{H^{-1}_\mu} 
\]
This estimates are the elliptic versions of the parametrix bounds. 
By Sobolev embeddings they imply bounds of the type of Theorem 4.

{\bf (iii) The low frequency parametrix.}
We first mollify the coefficients of $L_\mu$ on a scale $\mu^{-1/2}$ and note that
this does not affect the hypothesis of the Theorem.
We also modify its symbol for large $\xi$ and extend it to $\R^{2n}$ 
so that it is of size $ \mu^2$ and so that it satisfies 
\[ 
|\d^\alpha_x \d^\beta_{\xi} \tilde l_\mu (x,\xi)| \lesssim \left\{ \begin{array}{ll} 
\mu^{2-|\beta|} & \text{ if } |\alpha|\le 2 \\
\mu^{1+|\alpha|/2 - |\beta|} & \text{ if } |\alpha|\ge 3 
\end{array} \right.
\]  

By Theorem 3 of \cite{MR2094851} there exists a parametrix $T_{low}$
for $ \tilde l_\mu$ satisfying
\begin{equation}  \label{parametrix}
\begin{split} 
 \mu^{\frac1{n+1}} \Vert T_{low} f & \Vert_{L^{\frac{2(n+1)}{n+3}}} 
 + \mu^{1/2}  \Vert T_{low} f \Vert_{L^2}  
\\   & \lesssim
\inf_{f=f_1+f_2}  \mu^{-1/2} 
\Vert f_1 \Vert_{L^2} + 
\mu^{-\frac1{n+1}} \Vert f_2
\Vert_{L^{\frac{2(n+1)}{n-1}}} 
\end{split} 
\end{equation} 
and the error estimate
\begin{equation}  \label{parametrix1}
\begin{split} 
 \mu^{-1/2}
& \Vert (1-\tilde l_\mu^w(x,D) T_{low})  \psi(D/(2\mu R))  \phi f \Vert_{L^2} 
\\ & \lesssim   
\inf_{f=f_1+f_2} \mu^{-1/2} 
\Vert f_1 \Vert_{L^2} + 
\mu^{-\frac1{n+1}} \Vert f_2
\Vert_{L^{\frac{2(n+1)}{n-1}}} 
\end{split} 
\end{equation} 

{\bf (iv) The complete parametrix }
In the final step we combine the low and high frequency parametrices.
We set 
\[
T =   T_{high}(1-\psi(D/2\mu R) \phi_0 + \phi_0 
\psi(D/4\mu R)  T_{low}\psi(D/2\mu R) \phi_0  
\]
The estimate \eqref{T1} follows easily from the similar bounds for
$T_{high}$ and $T_{low}$. It remains to consider the error estimate.
We have
\begin{eqnarray*}
(I - LT )\phi f & =&  ( I - L T_{high}) (1-\psi(D/2\mu R) \phi f  \\ &+& 
 \phi_0  \psi(D/4\mu R)  (I - \tilde L_\mu^w T_{low})\psi(D/2\mu R)
 \phi f 
\\ &+& [ \tilde L_\mu^w, \phi_0  \psi(D/4\mu R)]
T_{low}\psi(D/2\mu R)
\phi f 
\\ &+& (L-\tilde L_\mu^w) \phi_0  \psi(D/4\mu R)
T_{low}\psi(D/2\mu R) \phi f
\end{eqnarray*}
For the first two terms we use the error estimates for $T_{high}$,
respectively $T_{low}$. In the third term the commutator has size $\mu$
in $L^2$ so we can use the $L^2$ bound for $T_{high}$. 
The operator
\[
(L-\tilde L_\mu^w) \phi_0  \psi(D/4\mu R)
\]
also has size $\mu$ in $L^2$ since the original coefficients differ 
from the mollified ones by $\mu^{-1}$. This complete the proof of the 
inequality \eqref{T2}.

{\bf Part B.}  We prove \eqref{Eu} by duality as in Section 3 of
\cite{MR2094851}.  Let $g \in W^{-\frac1{n+1},
  \frac{2(n+1)}{n+3}}_\mu$.  We decompose $\phi g$ as
\[ 
\phi g  =  h + L^*\overline{T} \phi g 
\]
where $\overline{T}$ is the operator of Theorem \ref{loc} constructed
for the formal adjoint operator $L^*$.  By part A of the theorem we have
\[ 
\begin{split}
(\delta \mu)^{-1/4} \mu^{-1/2} \Vert h \Vert_{L^2} 
 &+ (\delta \mu)^{1/4} \mu^{1/2} \Vert \overline{T}\phi g \Vert_{L^2} 
 +  \Vert \overline{T}\phi g \Vert_{W^{\frac1{n+1},\frac{2(n+1)}{n-1}}_\mu}
 \\ & \lesssim   
\Vert g \Vert_{W^{-\frac1{n+1}, \frac{2(n+1)}{n+3}}_\mu}.
\end{split} 
\]
Therefore we can write
\[ 
\begin{split}  
 | \langle \phi u,g \rangle|=& | \langle u, \phi g \rangle| 
\\  \le & 
| \langle u, h \rangle | + | \langle u,L^*\overline{T}\phi g \rangle |
\\  = & | \langle u, h \rangle | + | \langle Lu,\overline{T}\phi g \rangle |
\\ \lesssim & \!\!
\left( \!
(\delta \mu)^{1/4} \mu^{1/2} \Vert u \Vert_{L^2}\!
  + \! \inf_{Lu = f_1+f_2} (\delta \mu)^{-1/4} \mu^{-1/2} \Vert f_1  \Vert_{L^2}\!
  + \Vert f_2 \Vert_{W^{-\frac1{n+1}, \frac{2(n+1)}{n+3}}_\mu}
\! \right)   \\ &\times \Vert g \Vert_{W^{-\frac1{n+1}, \frac{2(n+1)}{n+3}}_\mu}.
\end{split}
\] 
This implies the estimate \eqref{Eu}.

{\bf Part C.} We begin with an $L^2$ estimate. 
 The principal  symbol 
of 
\[ \overline{L}_\mu = L_\mu   (\mu^2 + |D|^2)^{-1/2} \] 
 is 
\[ 
\bar{l}_\mu(x,\xi)=\Big(  - a^{ij}(x) \xi_i \xi_j + \mu^2 W(x) + 2 \mu g^j \xi_j\Big)
(\mu^2+ |\xi|^2)^{-1/2} . 
\]
A short calculation shows that 
\[
\begin{split}   
 \delta \mu  & - \{\Re \bar{l}_\mu(x,\xi),\Im  \bar{l}_\mu (x,\xi)  \}
 \lesssim |\bar{l}_\mu(x,\xi)|
\end{split}
\]
and hence, by Corollary II.14 of \cite{MR1944027}, we obtain the 
bound
\[
\delta \mu \| w\|_{L^2} \lesssim \| \overline{L}_\mu w\|_{L^2} +
\| w\|_{L^2}
\]
If $\delta \mu \gg 1$ then the norm of $u$ on the right hand side can
be hidden on the left hand side. Applying this to $w = (\mu^2 +
|D|^2)^{1/2} v$ we obtain
\begin{equation}  \label{l2c} 
\delta \mu \Vert v \Vert_{H^1_\mu}^2 
\lesssim  \Vert  L_\mu v \Vert_{L^2}^2   
\end{equation} 

For $u$ as in the theorem we write
\[
u = v + T L_\mu u
\]
The bounds for the second term come from part A.
On the other hand,
\[
L_\mu v = (1-L_\mu T) L_\mu u
\]
for which we can use the error estimate \eqref{T2} to obtain
\[ 
\begin{split}  
  (\delta \mu)^{-1/4} \mu^{-1/2} \Vert Lv \Vert_{L^2} \lesssim
  \inf_{L_\mu u = f_1+f_2}(\delta \mu)^{-1/4} \mu^{-1/2} \Vert f_1
  \Vert_{L^2} + \Vert f_2
  \Vert_{W^{-\frac1{n+1},\frac{2(n+1)}{n+3}}_\mu}
\end{split}
\] 
Then we successively apply \eqref{l2c} and  \eqref{Eu} to $v$,
concluding the proof. 
\end{proof}

\section{The $L^p$ Carleman inequality} 
\label{dispersive}

In this section we prove  Proposition \ref{carleman}.
We first conjugate with respect to the exponential weight. 
If we set $w= e^{h(\ln(|x|))} v$ then we can rewrite \eqref{cm}
in the form
\[
\begin{split} 
\Vert  w \Vert_{l^2W^{\frac1{n+1},\frac{2(n+1)}{n-1}}}  + 
\Vert \rho  w \Vert_{L^2}
\lesssim\inf_{L_h w = f_1+f_2}  \Vert\rho^{-1}    f_1 \Vert_{L^2} 
+ \Vert f_2 \Vert_{l^2 W^{-\frac1{n+1},\frac{2(n+1)}{n+3}}} 
\end{split}
\] 
where
\[
L_h = \Delta + V w + h'(\ln |x|)^2 |x|^{-2} -   h'(\ln |x|) \big[\nabla \frac{x}{|x|^2}
+ \frac{x}{|x|^2} \nabla\big]
\]
We
want to 
apply Theorem \ref{loc} on dyadic annuli 
\[ 
A_j = \{ x | 2^{j-1} < |x| < 2^{j+1}\}  
\]
The rescaling  $y= 2^{-j} x $ transforms this set to $A_0$ and the 
operator $L_h$ to 
\[
L_h^j = \Delta  + 2^{2j} \tilde V + h'(\ln (2^j |y|))^2 |y|^{-2}  - 
h'(\ln (2^j |y|)) \big[\nabla  \frac{y}{|y|^2}   + \frac{y}{|y|^2}  \nabla \big] 
\]

We verify that we can apply Theorem~\ref{loc} to $L^j_h$.  Since $h'$
varies slowly on the unit scale we can take the corresponding value
for $\mu$ to be
\[
\mu_j = \sqrt{2^{2j} + h^{'} (j \ln 2)^2}
\]
The coefficients $b$ and $c$ are given by
\[
 c = \mu_j^{-2} (2^{2j} V +
h'(\ln(2^j|y|))^2/|y|^2 )  , \qquad b_j = - \frac
{h'(\ln(2^j|y|))}{\mu_j}   \frac{y_j}{|y|^{2}}
\]
and are clearly of class $C^2$ and size $O(1)$. We have 
\[
\Re l_h^j(x,\xi) = -\xi^2 + c, \qquad \Im l_h^j(x,\xi) = 2 b \cdot \xi
\]
Their Poisson bracket has the form 
\[
\begin{split}
\{ -|\xi|^2 +  c, b\cdot \xi\}  =& \frac{h'(t)}{\mu |y|^2}  (-|\xi|^2 + c) + 2  y\cdot
  \xi \left (\frac{1}{|y|^4} - \frac{h''(t)}{h'(t)
      |y|^3}\right) b\cdot \xi 
\\ & - \frac{2^{2j} h'(t)}{ |y|^2\mu_j^3} y \cdot
\nabla V - \frac{2h'(t)^2 h''(t)}{|y|^{4} \mu_j^{3}}, \qquad t = \ln(2^j |y|)
\end{split}
\]
Then we can apply Theorem~\ref{loc} with $\delta$ comparable to the
size of the third term. For our choice of $h$ we have $|h''| \lesssim h'$ 
and also
\[
h''(t) < 0 \implies  h'(t) \ll e^{t}
\]
Hence we can choose
\[
\delta_j = \mu_j^{-3}\left(2^{2j} h'(j\ln 2) +h'(j\ln 2)^2 h''_+(j\ln 2)\right)
\]

Let $\phi \in C^\infty_0(\R)$ be a nonnegative function supported in 
$[-1,1]$ with 
\[ 
\sum_{j=-\infty}^\infty \phi^2(t-j) = 1 
\] 
and let $\phi_j(x) = \phi(\ln |x| -j)$.  After rescaling, part A of
Theorem~\ref{loc} yields a parametrix $T_j$ for $L_h$ in $A_j$ with the
property that

\[ 
\begin{split}  
 \Vert T_j g \Vert_{W^{\frac1{n+1}, \frac{2(n+1)}{n-1}}} 
 &+ \Vert   \rho    T_j g \Vert_{L^2} 
+
\Vert  \rho \frac{|x|}{h'(\ln |x|) +|x|}  \nabla ( T_j g) \Vert_{L^2} 
 \\  +  \Vert \rho^{-1}( L_h T_j - 1)& \phi_j g \Vert_{L^2}   
 \lesssim 
 \inf_{g=g_1+g_2} \Vert  \rho^{-1} g_1 \Vert_{L^2(A_j)}
 + \Vert g_2 \Vert_{W^{-\frac{1}{n+1},\frac{2(n+1)}{n+3}}(A_j)}.
\end{split}
\] 

We define a parametrix for $L_h$ by 
\[ 
T= \sum_{j=0}^\infty \phi_j T_j \phi_j 
\]
Summing up the bounds on $T_j$ we obtain a bound for $T$, 
\[
\begin{split} 
 \Vert T g  \Vert_{l^2W^{\frac{1}{n+1},{\frac{2(n+1)}{n-1}}}} 
  +  \Vert  \rho    T g  \Vert_{L^2} 
 \lesssim  \inf_{g=g_1+g_2} \Vert  \rho^{-1}  g_1 \Vert_{L^2} + 
\Vert g_2 \Vert_{l^2W^{-\frac1{n+1},\frac{2(n+1)}{n+3}}}.
\end{split}
\] 
We also compute the error
\[
1-L_h T = \sum_{j=0}^\infty  \phi_j(1-L_h T_j) \phi_j - \sum_{j=0}^\infty
[L_h,\phi_j] T_j \phi_j
\]
Since 
\[
[L_h,\phi_j] = O(|x|^{-1}) \nabla + O(  h'(\ln |x|)|x|^{-2})
\]
and
\[
|x|^{-1} \lesssim \rho^2 \frac{|x|}{ h'(\ln |x|) +|x|}, \qquad  h'(\ln |x|) |x|^{-2}
\lesssim \rho^2
\]
we can bound the error by
\[
\|\rho^{-1} (1-LT) g\|_{L^2} \lesssim  \inf_{g=g_1+g_2} \Vert  \rho^{-1}  g_1 \Vert_{L^2} + 
\Vert g_2 \Vert_{l^2W^{-\frac1{n+1},\frac{2(n+1)}{n+3}}}.
\]

Now, after the construction of the parametrix the assertion of
Proposition \ref{carleman} follows exactly as the corresponding part
of Theorem \ref{loc}.  We repeat the argument.  Split $w$ into
\[
 w = v + TLw  
\]
Then the second term satisfies the desired bounds while for the
first we know that 
\[ 
 \Vert \rho^{-1}   L v \Vert_{L^2}  = 
\Vert   \rho^{-1}  (LT-1) Lw \Vert_{L^2} 
  \lesssim       \inf_{  Lw =g_1+g_2} \Vert  \rho^{-1}  g_1 \Vert_{L^2} + 
\Vert g_2 \Vert_{l^2W^{-\frac1{n+1},\frac{2(n+1)}{n+3}}}. 
\]
Lemma \ref{l2}  allows us to also estimate
\[
\| \rho v\|_{L^2}
\]

On the other hand by  Theorem  \ref{loc}, B rescaled and applied to $v$ in $A_j$ we get
\[
\Vert \phi_j v \Vert_{W^{\frac1{n+1}, \frac{2(n+1)}{n-1}} } 
 \lesssim \Vert \rho v \Vert_{L^2(A_j)} +   \Vert  \rho^{-1}      Lv \Vert_{L^2(A_j)} 
\]
and after summation in $j$,
\[
 \Vert v \Vert_{l^2W^{\frac1{n+1}, \frac{2(n+1)}{n-1}} } 
 \lesssim \Vert \rho v \Vert_{L^2} +   \Vert  \rho^{-1}      Lv \Vert_{L^2} 
 \]
thereby concluding the proof.

\section{ Equations with gradient potentials }

In this section we discuss the corresponding results which are
obtained when short range gradient potentials are added.
Thus we consider equations of the form
\begin{equation}
 \label{fund1}  
(- \Delta  - V ) u = W u  + Z^l \nabla u +  \nabla Z^r  u
\end{equation}
with $V$ and $W$ as before. The gradient potential $Z=(Z^l,Z^r)$ is 
subject to the following conditions:

\begin{assumption}[The short range gradient potential] 
 The gradient potential $Z \in l^\infty(L^n)$ satisfies
\begin{equation}\label{zln}  
\limsup_{j\to \infty} \|Z\|_{L^n(\{ x| 2^j \le |x| \le 2^{j+1}\})}
\leq \delta
\end{equation}
In addition for some $R \gg \| V\|_{L^\infty}$ the low 
frequency part $S_{<R}Z$ of $Z$ satisfies the conditions 
in Assumption A2.
\end{assumption}

The $L^n$ assumption is natural due to scaling. The low frequency 
condition is also natural, since on the characteristic set of $-\Delta-V$ 
the frequency has size $O(1)$, and at frequency one there is no
difference between the potential and the gradient potential.
Under these conditions we have

\begin{theorem} \label{uniquenessz} Assume that  $V$, 
  $W$ and $Z$ satisfy Assumptions A1,A2 respectively A3. Let
  $\tau_1>\tau_0$ and assume that $\delta$ is sufficiently small. Let
  $u \in H^1_{loc}(\R^n)$ satisfy \eqref{fund} and
  $(1+|x|^2)^{\tau_1-\frac12} u \in L^2$.  Then $u\equiv 0$.
\end{theorem} 

By scaling we obtain the following result on the absence of embedded
eigenvalues:

\begin{corollary}
  Assume that $V$, $W$ and $Z$ satisfy Assumptions A1,A2 respectively
  A3 with $\delta = 0$. Then there are no embedded eigenvalues for
the operator 
\[
-\Delta -W - Z^l \nabla  - \nabla Z^r  
\]
\end{corollary}

The problem of introducing gradient potentials has long been
considered in the context of the unique continuation and the strong
unique continuation problems for the same operators as here.  There
the key breakthrough came in Wolff's work \cite{MR96c:35068} who proved that $Z
\in L^n$ suffices for the unique continuation property. He also
obtained the same result for strong unique continuation but only in
low dimension. Later his ideas were used by the authors in \cite{MR2001m:35075} to
complete the picture for strong unique continuation in high dimension,
working with gradient potentials $Z \in l^1L^n$ . This latter paper is
more relevant to the present context as it provides Carleman estimates
in largely the same format as here.

Ideally, one would like to include matching gradient estimates to our
$L^p$ Carleman inequalities. This would solve the problem but
unfortunately cannot work. Wolff's contribution was to show that by
osculating the weight one can considerably improve the bounds for the
gradient term in the equation.  Thus the choice of weights ultimately
depends both on the gradient potentials and on the solution $u$. In
our context this argument is needed only at spatial scales where the
frequency of the conjugated operator is larger than one. Elsewhere the
gradient does not contribute much to the problem. Thus we are led to
consider perturbed weights
\begin{equation}
\psi_{\e,\tau} (x) = h_\epsilon(\ln|x|) + k(x)
\label{kc0} \end{equation} 
where $k$ is not spherically symmetric but is small in an appropriate
sense. The assumptions on $k$ are summarized in what follows:

\begin{equation}
\left\{ 
\begin{array}{l}
\text{supp}\ k \subset \{|x| \leq \tau^2\} \cr \cr
|x|^\alpha |\nabla^\alpha k(x)| \ll h'(\ln|x|) \quad \alpha = 1,2,3
\end{array}
\right.
\label{kc}\end{equation}

Part of the Carleman estimates below describes what happens in
elliptic regions of the conjugated operator $L_\phi$. To select
(part of) this elliptic region we introduce a pseudodifferential
operator $\chi_{>R}$ which selects the region 
\[
E = \{ |x| \gtrsim \tau^2,\ |\xi| \gtrsim R\}
\]
Here both the truncation in $x$ and in $\xi$ are done on the dyadic
scale, while $R$ is chosen sufficiently large so that $E$ is away from
the characteristic set of $P_\phi$.  Then the Carleman estimates are
as follows:

\begin{theorem}
  Assume that the long range potential $V$ satisfies A1. Let $Z$
  satisfy A3 with
\[
\|Z\|_{l^\infty L^n} + \|S_{<R} Z\|_{l^\infty X+ \langle x
  \rangle^{-1} L^\infty} \leq 1 
\]
Then for each $0 < \epsilon \leq \epsilon_0$, $\tau$ large enough,
$\tau_1 > \tau_0$ and $v$ which satisfies $(1+|x|^2)^{\tau_1-\frac12} v \in
L^2$ there is a weight perturbation $k$ satisfying \eqref{kc0},\eqref{kc} so that
the following estimate holds with constants independent of $0 <
\epsilon \leq \epsilon_0$, $\tau > \tau_0$:
\begin{equation} \label{cmz}
\begin{split}
& \Vert e^{\psi_{\e,\tau} (x)}  v \Vert_{l^2W^{\frac1{n+1}, \frac{2(n+1)}{n-1}}} 
+  \Vert \chi_{>R} e^{\psi_{\e,\tau} (x)}  v \Vert_{H^1}
+  \Vert e^{\psi_{\e,\tau} (x)}  \rho  v \Vert_{L^2} \\ + &
\| e^{\psi_{\e,\tau} (x)} Z^l \nabla v\|_{
  l^2W^{-\frac{1}{n+1},\frac{2(n+1)}{n+3}}+
\chi_{>R} H^{-1}}+\| e^{\psi_{\e,\tau} (x)}  \nabla Z^r v\|_{
  l^2W^{-\frac{1}{n+1},\frac{2(n+1)}{n+3}}+
\chi_{>R} H^{-1}}
\\ \lesssim   & \inf_{f_1+f_2=(-\Delta - V)v }  \Vert 
e^{\psi_{\e,\tau} (x)} \rho^{-1}   f_1 \Vert_{L^2} + \Vert e^{\psi_{\e,\tau} (x)}
f_2 \Vert_{l^2W^{-\frac{1}{n+1},\frac{2(n+1)}{n+3}}+\chi_{>R} H^{-1}}
\end{split}
\end{equation}
\end{theorem}

The key feature of the theorem is that the weight $\psi_{\e,\tau} (x)$
depends both on the potential $Z$ and on the solution $v$ itself.
Once this result is established, it leads as before to the conclusion
that solutions to \eqref{fund1} must be compactly supported. Then (a
variation of) Wolff's weak unique continuation result
\cite{MR96c:35068} takes over and implies that $v$ must be identically
$0$. We also refer the reader to \cite{MR2001m:35075}, where the estimates are
formulated in a way similar to this paper, and where both left and
right gradient potentials are considered.

The proof requires the following steps:

(i) Conjugate the equation with respect to the exponential weight and
set $w=e^{\psi_{\epsilon,\tau}} v$.  This elliminates the exponential
weight from the equation and replaces the operator $-\Delta - V$ by
its conjugated operator $L_{\psi_{\epsilon,\tau}}$.

(ii) prove the $L^2$ estimate for $v$ uniformly for all weights
$\phi_{\e,\tau}$ with $k$ satisfying \eqref{kc}. This is done exactly
as in Section~\ref{l2sect}. The size of the perturbation $k$ is so
that its effect is negligible in this computation.

(iii) prove the estimate \eqref{cm}, again uniformly with
respect to all choices for $k$. This repeats the arguments in
Section~\ref{dispersive} with no change.

(iv) Add in the $H^{-1}$ and $H^1$ norms, thus proving \eqref{cmz} for
$Z=0$. This is done in an elliptic fashion, by constructing an
elliptic parametrix for $L_\phi$ away from its characteristic set.
For this the norms involving $\rho$ are used only to estimate errors,
while the $L^p$ norms are all used via $L^2$ norms and Sobolev
embedding. The standard pseudodifferential calculus can be applied
since the coefficients of $P_\phi$ are smooth on the dyadic scale in
$x$.

(v) observe that the $L^2$ estimates allow localization on the
dyadic spatial scale. Thus we separate the estimate into two regions,
$\{ |x| < \tau^2\}$ and $\{|x| > \tau^2/2\}$.

(iv) show that within the first region it is possible to choose the
weight $k$ so that the estimate with $Z$ included holds. This is the
part that uses Wolff's osculation lemma, and it is explained in detail
in \cite{MR2001m:35075}. Our case here is somewhat simpler than in
\cite{MR2001m:35075} since in the region $ \{ |x| < \tau^2\}$ we have
uniform convexity of the weight, $h'' \approx h'$. Also the $L^p$
bound here is stronger than in \cite{MR2001m:35075}, which only makes things
better.

(v) prove the estimate in each dyadic component of the second region
$\{|x| > \tau^2/2\}$ with $Z$ included.  This starts from the
estimates without $Z$ and uses only elliptic bounds. We outline the
argument. Since we use dual norms on the left and on the right of
\eqref{cmz}, it suffices to do it for the $Z \nabla$ term. The bounds
for $\nabla Z$ will work out similarly but in dual spaces.

 We split $Z$ into a low and a high frequency part,
\[
Z = Z_{<R} + Z_{> R}
\]
and the gradient also,
\[
\nabla = \nabla_{<R/2} + \nabla_{>R/2}
\]

Using the $L^{\frac{n+1}2}$ bound on $Z_{<R/2}$ we can directly estimate
the contribution of $Z_{<R} \nabla_{<R/2}$ 
which is  located at low frequency.

The contribution of $Z_{>R} \nabla_{<R/2}$ lies at  high frequency, so
it suffices to bound it in $H^{-1}$. We can actually bound it in $L^2$,
\[
\| Z_{>R} \nabla_{<R/2} w\|_{L^2} \lesssim \|Z\|_{L^n} 
\|w\|_{W^{\frac1{n+1}, \frac{2(n+1)}{n-1}}}
\]
For $Z \nabla_{>R/2}$ we can use the $H^1$ bound to write
\[
\| Z \nabla_{>R/2} w\|_{L^{\frac{2n}{n+2}}} \leq \|Z\|_{L^n}
  \|\chi_{>R/2} w\|_{H^1}
\]
and conclude by Sobolev embeddings.

\section{ Asymptotically flat metrics }

In this section we describe how the results on the absence of
embedded eigenvalues extend to variable coefficient asymptotically
flat metrics. We replace the Laplacian with a second order elliptic
selfadjoint operator 
\[
L = -  \d_j a^{jk}  \partial_k + i(b^j
\partial_j+\partial_j b^j) + c
\]
where the coefficients $a,b,c$ are real. 
We assume that $P$ is flat at infinity in the sense that (see Definition
\ref{c2x}) :
\begin{equation}
\begin{split}
&a^{jk} ,b^j,c \in C^2_{\langle x\rangle}  \subset  L^\infty
\\
& \limsup_{|x| \to \infty} |x|  |\nabla a^{ij}| \leq \delta_0,
\\ & 
 \limsup_{|x| \to \infty} |b(x)| +  |x|  |\nabla b(x)|  \leq \delta_1,\qquad
   \limsup_{|x| \to \infty} |c(x)|  +  |x|  |\nabla c(x)|  \leq \delta_1^2
\end{split}\label{coef}\end{equation}

 We also slightly strenghten the assumption A3 to make it stable with
 respect to changes of variable:

\begin{assumption}[The short range gradient potential] 
 The gradient potential $Z \in l^\infty(L^n)$ satisfies
\begin{equation}\label{zlna}  
\limsup_{j\to \infty} \|Z\|_{L^n(\{ x| 2^j \le |x| \le 2^{j+1}\})}
\leq \delta
\end{equation}
In addition $\langle D \rangle^{-N} Z$ satisfies the conditions 
in Assumption A2 for some $N$ sufficiently large.
\end{assumption}
Then we have 

\begin{theorem} \label{uniquenessvc} 
  Assume that $W$, $V$ and $Z$ satisfy Assumptions A1,A2 and A4 with
  small enough $\delta$, that $\tau_1 > \tau_0$ and that the
  coefficients of $P$ satisfy \eqref{coef} with $\delta_0$ and
  $\delta_1$ sufficiently small.  If $u \in H^1_{loc}(\R^n)$ solves
\begin{equation}
 L u + Vu  = W u + Z^l \nabla u  + \nabla Z^r  u
\end{equation}
and $(1+|x|^2)^{\tau_1-\frac12} u \in L^2$  then $u\equiv 0$.
\end{theorem} 

The assumption of Theorem \ref{uniquenessvc} are not scale invariant. 
For the following  straightforward consequence we rescale the operator. 

\begin{corollary}
  Assume that the coefficients of the operator $P$ satisfy
  \eqref{coef} with $\delta_0$ sufficiently small. Let $W$, $Z$ be as
  in Assumptions A2, A4 with $\delta = 0$.  Then there exists $C > 0$
  so that $P+W$ has no eigenvalues $\lambda > C\delta_1$.
\end{corollary}

The proof  follows the same outline as in the constant coefficient
case. We describe the steps in what follows, and discuss the necessary
modifications. 

First one needs to augment \eqref{coef} to gain also the relation
\begin{equation}
 \limsup_{|x| \to \infty} |a(x)-I_n|  \lesssim \delta_0
\label{coef1} \end{equation}
This is achieved using a change of coordinates somewhat similar to the
one introduced in \cite{MR2001m:35075}. Due to \eqref{coef},
within each spatial dyadic region this can be achieved with a linear 
change of coordinates. But from one dyadic region to the next
these linear maps differ by $O(\delta)$. Hence gluing them together
yields a nonlinear function $\chi$ which achieves \eqref{coef1} and
has the regularity
\[
|\partial^\alpha \chi(x)| \lesssim \delta_1 |x|^{1-|\alpha|} \qquad
|\alpha| \geq 2
\]
It is easy to verify that such a change of coordinates does not affect
$\delta_1$ by more than a fixed  factor.

If $\chi$ were linear then the Assumption A1 on $V$ would rest
unchanged. As it is, we have to modify $\tau_1$ by $O(\delta_0)$,
which is suitably small.

Finally, the operator $L$ is still $L^2$ selfadjoint in the new coordinates
but with respect to the measure given by the Jacobian $J$ of the 
change of coordinates. This implies that $JL$ is selfadjoint with 
respect to the Lebesque measure. This requires replacing $V$ and $W$
by $JV$ and $JW$, which has no significant effect on our assumptions.

Once \eqref{coef1} is gained the Carleman estimates \eqref{cm} in
Proposition~\ref{carleman} remain valid with essentially no change.
The only minor modification that is needed is concerned with what
happens within a compact set, where we have no control over the
geometry of the coefficients $a^{ij}$ in the principal part. But this
can be easily addressed by adding some additional convexity to the
exponential weight within this compact set. Precisely, a weight of the
form
\[
h(t) = \tau e^{\lambda t} 
\]
would suffice for bounded $t$ provided $\lambda$ is large enough.

The $L^2$ Carleman estimates are
 established using integration by parts, and do not require any bounds
on the second derivatives of $a^{ij}$.

The $L^p$ Carleman estimates are derived from the $L^2$ ones exactly
as in Section~\ref{dispersive}. For comparison purposes, we recall
that the $L^p$ estimates proved in \cite{MR2094851} and \cite{MR2001m:35075}
only require bounds on the first derivatives of $a^{ij}$. This is
because the spatial localization which is allowed by the Carleman
estimates is on a scale on which one is allowed to freeze the $a^{ij}$
modulo negligible errors.  The same applies here for $|x| \lesssim
\tau^2$ (which corresponds to $e^t \lesssim \tau^2$). However, beyond this
threshold the rescaled skewadjoint part becomes very small and the
problem is close to the spectral projection estimates respectively the
Strichartz estimates for wave equations with $C^2$ coefficients. The
spatial localization scale is $h'(\ln(|x|))^{-\frac12} |x|$ while the
frequency, instead of decaying, remains $O(1)$ due to the long range
potential $V$.  Hence the difference between $P$ and its frozen
coefficient version is $O(h'(\ln(|x|))^{-\frac12})$, which is more than the
constant $\rho^2$ in the $L^2$ estimates. This is why we need 
also bounds on the second derivatives of $a^{ij}$, as required by
pTheorem~\ref{loc}.

Finally, the gradient potential can be added in as explained in the
previous section.
\section{Appendix}

We consider a dyadic partition of unity in $\R^n$,
\[
1 = \sum_{j=-\infty}^\infty \chi_j(x)
\]
where $\chi_j(x)=\chi_0(2^{-j} x)$ is supported in $|x| \approx 2^j$. 
 We also consider bump functions $\tchi_j(x)= \tchi_j(2^{-j}x)$
 with slightly larger support, which equal $1$
within the support of $\chi_j$ such that $\tchi_j \tchi_l=0$ if $|j-l|\ge 2$.  
\begin{lemma}
Let $1 < p < \infty$ and $ -\frac{n}{p'} <  s < \frac{n}p  $. Then
\[
\|u\|_{W^{s,p}}^p \approx \sum \| \chi_j u\|_{W^{s,p}}^p 
\]
\end{lemma}
\begin{proof}
Let $(u_j)$ be a sequence in $W^{s,p}$. 
Arguing by duality it suffices to prove the bound
\[
\| \sum_{j=-\infty}^\infty \chi_j u_j\|_{W^{s,p}}^p \lesssim \sum \| u_j \|_{W^{s,p}}^p 
\]
 With $\langle D \rangle= (1+|D|^2)^{1/2}$ we have
\[
\| \sum_{j=-\infty}^\infty \chi_j u_j\|_{W^{s,p}} = \Vert \langle D \rangle^{s} \sum \chi_j u_j\|_{L^p}
\]
We write
\[
\langle D \rangle^{s} \sum_{j=-\infty}^\infty \chi_j u_j = 
\sum_{j=-\infty}^\infty \tchi_j \langle D \rangle^s \chi_j u_j
+  \sum_{j=-\infty}^\infty (1-\tchi_j) \langle D \rangle^s \chi_j u_j
\]
The terms in the first sum have almost disjoint supports and are easy to
estimate. It remains to consider the second sum. We use  bounds
on the kernel of $\langle D \rangle^{-s}$ and its derivatives 
to estimate 
\[
 |(1-\tchi_j) \langle D \rangle^{s} \chi_j u_j(x) | \lesssim   \|u_j\|_{L^p} 2^{(s+\frac{n}{p'})j} (2^{j} +
 |x|)^{-n-s}. 
\]
Then we conclude using 
\[
\| \sum_{j=-\infty}^\infty a_j 2^{(s+\frac{n}{p'})j} (2^{j} +
 |x|)^{-n-s} \|_{L^p}^p \approx 
\sum_{j=-\infty}^\infty |a_j|^p \qquad s > -\frac{n}{p'}
\]
\end{proof}

This is the main ingredient in the proof of

\begin{proposition}\label{small} 
Let $\delta > 0$. Suppose that $W \in X$ (see Definition \ref{X}).
Then we have
\[
\lim_{\alpha \to 0} \|W\|_{ X(B(0,\alpha))}  =0
\]
\end{proposition}

\begin{proof}
  We assume that $n \geq 3$, the case $n=2$ is similar. The result
  follows from the estimate
\begin{equation}
\|W\|_X \approx  \| \chi_j W\|_{l^\frac{n+1}2 (X)}
\end{equation}
For one direction  we write 
\begin{eqnarray*}
|\langle W u,v \rangle| &=& |\sum  \langle \chi_j W \tchi_j u, \tchi_j v \rangle|
\\ &\lesssim& \sum \| \chi_j W\|_{X} \|\tchi_j
u\|_{W^{\frac1{n+1},\frac{2(n+1)}{n-1}}} 
\|\tchi_j v\|_{W^{\frac1{n+1},\frac{2(n+1)}{n-1}}}
\\ &\lesssim&  \| \chi_j W\|_{l^\frac{n+1}2} \| \tchi_j
u\|_{l^\frac{2(n+1)}{n-1}W^{\frac1{n+1},\frac{2(n+1)}{n-1}}} \|  \tchi_j v\|_{l^\frac{2(n+1)}{n-1}W^{\frac1{n+1},\frac{2(n+1)}{n-1}}}
\\ &\lesssim&  \| \chi_j W\|_{l^\frac{n+1}2}
\|u\|_{W^{\frac1{n+1},\frac{2(n+1)}{n-1}}}
\|v\|_{W^{\frac1{n+1},\frac{2(n+1)}{n-1}}}
\end{eqnarray*}
For the other, we consider separately sums with $j$ even and with $j$ odd:
\begin{eqnarray*}
 \sum_{j \ even}  \langle \chi_j W  u_j,  v_j \rangle &=& 
 \sum_{j \ even}   \langle \chi_j W \tchi_j u_j, \tchi_j v_j \rangle 
\\ &=& \langle W \sum_{j \ even}  \tchi_j u_j, \sum \tchi_j v_j \rangle 
\\ &\lesssim& \|W\|_X \|\sum_{j \ even} \tchi_j u_j\|_{ W^{\frac1{n+1},\frac{2(n+1)}{n-1}}}
 \|\sum_{j \ even} \tchi_j v_j\|_{ W^{\frac1{n+1},\frac{2(n+1)}{n-1}}}
\\ &\lesssim& \|W\|_X \|u_j\|_{ l^\frac{2(n+1)}{n-1} W^{\frac1{n+1},\frac{2(n+1)}{n-1}}}
\|u_j\|_{ l^\frac{2(n+1)}{n-1} W^{\frac1{n+1},\frac{2(n+1)}{n-1}}}
\end{eqnarray*}

\end{proof}

 \end{document}